     \documentclass[12pt]{article}   
     \usepackage{epsfig}
     \newlength{\dinwidth}                       
     \newlength{\dinmargin}                      
\newcommand\hepph[1]{hep-ph/#1}

\newcommand\hepex[1]{hep-ex/#1}

\newcommand\jhep[3]{{{\it J. High Energy Phys. }{\bf #1} (#2) #3}}

\newcommand\npb[3]{{{\it Nucl. Phys. }{\bf B #1} (#2) #3}}

\newcommand\plb[3]{{{\it Phys. Lett. }{\bf B #1} (#2) #3}}

\newcommand\prep[3]{{{\it Phys. Rep. }{\bf #1} (#2) #3}}

\newcommand\zpc[3]{{{\it Z. Physik }{\bf C #1} (#2) #3}}

\newcommand\jetp[3]{{{\it Sov. Phys. JETP }{\bf #1} (#2) #3}}

\newcommand\jetpl[3]{{{\it JETP Lett. }{\bf #1} (#2) #3}}

\newcommand\epjc[3]{{{\it Eur. Phys. J. }{\bf C #1} (#2) #3}}

\def\ga{\gamma}

\def\lsim{\mathrel{\rlap{\lower4pt\hbox{\hskip1pt$\sim$}}
    \raise1pt\hbox{$<$}}}                
\def\gsim{\mathrel{\rlap{\lower4pt\hbox{\hskip1pt$\sim$}}
    \raise1pt\hbox{$>$}}}                
\begin{document}
\begin{flushright}
  Bicocca--FT--99--14\\
  hep-ph/9906302\\
  May 1998
\end{flushright}

\vspace*{10mm}
\begin{center}  \begin{Large} \begin{bf}
Small-$x$ final states and the CCFM equation\footnote{Work supported in
  part by the E.U. QCDNET 
 contract FMRX-CT98-0194}\footnote{Contribution to workshop on Monte
 Carlo generators for HERA physics.}\\ 
  \end{bf}  \end{Large}
  \vspace*{5mm}
  \begin{large}
G.P. Salam$^a$\\
  \end{large}
\end{center}
$^a$ INFN --- Sezione di Milano, and Universit\`a di Milano--Bicocca,
Milano, Italy\\
\begin{quotation}
\noindent
{\bf Abstract:} The status of the Milan group's work on CCFM-based
phenomenology at small $x$, and possible directions of future
investigation, are discussed.
\end{quotation}

In order to study final states in QCD it is generally necessary to
take into account the coherence of QCD radiation \cite{coherence}. At
small $x$, coherence is embodied in the CCFM equation \cite{CCFM}.  In
\cite{BMSS} a first analysis of the CCFM equation was carried out with
regard to small-$x$ phenomenology at HERA. The information gained was
to have served as a basis for a CCFM-based Monte Carlo event
generator.  The main results are summarised here, together with some
discussion of possible future work.

As a first step, we studied the freedom available in implementing the
CCFM equation, and noted that the non-Sudakov form factor (which
resums virtual corrections at small $x$) is not uniquely defined.
There are two possible choices for it. One leads to small NLL
corrections compared to BFKL \cite{BFKL,NLL}, the other to large NLL
corrections (much bigger that the true NLL corrections to BFKL).

Soft emissions (those from the $1/(1-z)$ part of the splitting
function) were neglected throughout the analysis.  Formally this
should have been acceptable for the observables we considered. Their
inclusion would have been another source of NLL corrections.

As a phenomenological constraint on the free parameters of our
evolution we fitted the small-$x$ $F_2$ structure function \cite{F2}.
It was found that of the two possible implementations of the
non-Sudakov form factor, only the one with the large corrections was
able to reproduce the $F_2$ data. We then examined other, more
exclusive, observables such as the transverse energy flow \cite{Et},
charged-particle transverse-momentum spectra \cite{pt} and the
forward-jet cross section \cite{fjZS,fjH1}. No hadronisation was
implemented. Nor did we include the contribution to the final state
from the quark box.

\begin{figure}[htbp]
  \begin{center}
    \epsfig{file=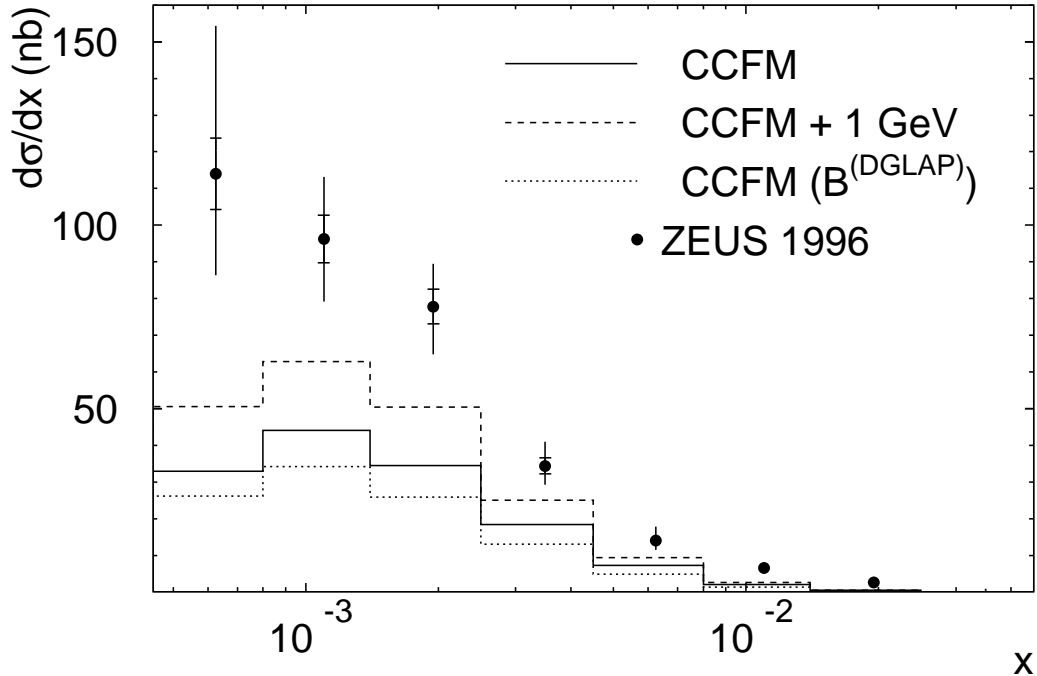,width=0.8\textwidth}
    \caption{Forward-jet cross section from \cite{BMSS} as compared to
      ZEUS data \cite{fjZS}. The differences between the three
      histograms are indicative of the uncertainty on the
      theoretical result.\vspace{-0.5cm}}
    \label{fig:fj}
  \end{center}
\end{figure}

All final-state observables were systematically low (especially the
$E_t$ flow and the forward-jet cross section, which were off by a
factor of $2$). In the case of the $E_t$ flow this might be
explainable by the absence of the contributions from soft emissions
and hadronisation corrections.  Less so for the other observables. As
an example, figure~\ref{fig:fj} shows our results for the forward-jet
cross section, together with an indication of the uncertainties, as
compared to the data.

What is our understanding as to why we failed to reproduce the data?
We suspect that the choice of implementation of the non-Sudakov form
factor is largely to blame. We needed a `source of subleading
corrections' in order to be able to fit $F_2$. This was given to us by
our modification of the non-Sudakov form factor. But we subsequently
discovered that this modification leads to strange behaviours, such as
small-$x$ cross sections (e.g.\ for forward jets) which first decrease
as $x$ decreases, and only at very small $x$ start to increase: our
source of subleading corrections was too large and had the wrong
characteristics.

Does this mean that the CCFM equation is inconsistent with the data?
Not necessarily.  It should be remembered that we did not include the
soft emissions. It is quite possible that they alone would have been a
sufficient source of subleading corrections, in order to allow a fit
to the structure function data, without requiring a modification of
the non-Sudakov form factor (the soft emissions lead to somewhat
smaller, but still sizeable NLL corrections). They would also have
increased activity in the final state. Such an approach might
therefore have led to much better agreement with the data.  (This
would have been close in physics content to the SMALLX \cite{SMALLX}
program, which is currently being studied by Hannes Jung \cite{Jung},
who finds that it does actually reproduce a range of data relatively
well).

We have not however taken the seemingly logical step of implementing
the soft emissions within our approach. Why? Two practical issues are
involved. One is undoubtedly the difficulty involved --- the inclusion
of soft emissions complicates enormously the numerical solution of the
CCFM equation.\footnote{It should be pointed out that the recent
  results on the LL equivalence of CCFM (including soft emissions) and
  BFKL final states \cite{gps99} is not sufficient to allow one to use
  the BFKL equation as the basis of a Monte Carlo --- indeed as
  discussed in \cite{dis99}, at subleading orders, in DIS, the BFKL
  equation has serious pathologies in its final states.}
Additionally, as mentioned above, there is already a program in
existence, SMALLX, which implements CCFM including soft emissions.
There, the difficulty of implementing soft emissions is traded for
difficulties of use (it is a forward-evolution Monte Carlo generating
weighted events, with considerable fluctuations in the weights), and
restricted flexibility in certain areas such as the inclusion of the
appropriate scale for the running coupling.

But there are also physics issues of relevance. In the past year the
NLL BFKL corrections have become available \cite{NLL} and there has
been significant progress in understanding the physical origins of the
numerically dominant parts of the corrections \cite{NLLresum}. The
CCFM equation (even with soft emissions and running coupling) is
missing two of the main physical parts of the NLL corrections. There
is a contribution related to the finite part of the gluon splitting
function (while the $z\to0$ and $z\to1$ singular parts are included by
the CCFM equation, the part which is finite at both $z=0$ and $z=1$ is
not). This might be quite straightforwardly included in the CCFM
equation. But there is also a component of the NLL corrections which
ensures the symmetry of the evolution --- the requirement that
evolution from a small (transverse) scale to a large scale give the
same answer as evolution from a large scale to a small scale.  The
CCFM equation does not satisfy this symmetry, corresponding
technically to the absence of the component of the NLL corrections
which goes as $1/(1-\ga)^3$ ($\ga$ is the Mellin transform variable
conjugate to the transverse momentum). Nor is it immediately clear how
to introduce this symmetry without modifying the whole structure of
the CCFM equation. (Such a modification has been proposed, in the form
of the LDC model \cite{LDC}, but it changes the small-$x$ leading
logarithms).

Given that without these contributions, in terms of the physics
included, it would be difficult to improve significantly over the
already-existing SMALLX program, we have decided to place on hold our
plans for the construction of a small-$x$ Monte Carlo event generator.
Such a project is of considerable importance. But an important (and
ongoing) part of it is to understand how to include the various
physical components of the NLL corrections outlined above, into an
equation such as the CCFM equation which is suitable for final-state
studies.

\section*{Acknowledgements}
The work on implementing a CCFM description of final states was done
in collaboration with Giulio Bottazzi, Giuseppe Marchesini and Massimo
Scorletti. I would like also to thank Yuri Dokshitzer, Marcello
Ciafaloni and Dimitri Colferai for discussions.

\end{document}